# Pain relief devoid of opioid side effects following central action of a silylated neurotensin analog

Pascal Tétreault[a,b,c,d,1], Élie Besserer-Offroy[a,c,d,1,2], Rebecca L Brouillette[a,c,d], Adeline René[e], Alexandre Murza[a,c,d], Roberto Fanelli[e], Karyn Kirby[a,c,d], Alexandre J Parent[a,c,d], Isabelle Dubuc[f], Nicolas Beaudet[b], Jérôme Côté[a,c,d], Jean-Michel Longpré[a,c,d], Jean Martinez[e], Florine Cavelier[e,3,*], Philippe Sarret[a,b,c,d,3,*]

[a] Department of Pharmacology-Physiology, Faculty of Medicine and Health Sciences, Université de Sherbrooke, 3001 12th Avenue North, Sherbrooke, Québec, J1H5N4, Canada.

[b] Department of Anesthesiology, Faculty of Medicine and Health Sciences, Université de Sherbrooke, 3001 12th Avenue North, Sherbrooke, Québec, J1H5N4, Canada.

[c] Institut de pharmacologie de Sherbrooke, Université de Sherbrooke, 3001 12th Avenue North, Sherbrooke, Québec, J1H 5N4, Canada.

[d] Centre de recherche du Centre hospitalier universitaire de Sherbrooke, CIUSSS de l'Estrie – CHUS, 3001 12th Avenue North, Sherbrooke, Québec, J1H5N4, Canada

[e] Institut des Biomolécules Max Mousseron, UMR-5247, CNRS, Université Montpellier, ENSCM, 19 Place E. Bataillon 34095 Montpellier cedex 5, France.

[f] Department of Pharmacy, Faculty of Medicine and Pharmacy, Université de Rouen, 1 rue Thomas Becket, 76821 Mont-Saint-Aignan cedex, France.

[1] These authors contributed equally to this work

[2] Present address: Department of Pharmacology and Therapeutics, McGill University, Montréal, Québec, Canada.

[3] Lead Authors

**Email addresses of authors:**

Pascal.Tétreault@USherbrooke.ca (PT)

Elie.Besserer@McGill.ca (ÉBO)

Florine.Cavelier@umontpellier.fr (FC)

Philippe.Sarret@USherbrooke.ca (PS)

* **Co-Corresponding Authors:**

**Philippe Sarret, Ph.D.**
Dept of Pharmacology-Physiology
Université de Sherbrooke
3001, 12e Avenue Nord
Sherbrooke, QC, Canada, J1H 5N4
Philippe.Sarret@USherbrooke.ca
Phone: +1 (819) 821-8000 ext. 72554

**Florine Cavelier, Ph.D.**
IBMM, UMR-CNRS-5247
Université de Montpellier
19 Place Eugène Bataillon
34095 Montpellier Cedex 5, France
Florine.Cavelier@umontpellier.fr
Phone: +33 (0) 467 14 37 65

**Chemical compounds studied in this article**

Neurotensin (PubChem CID 25077406)

PD149163 (PubChem CID 73064239)

**Abstract**

Neurotensin (NT) exerts naloxone-insensitive antinociceptive action through its binding to both $NTS_1$ and $NTS_2$ receptors and NT analogs provide stronger pain relief than morphine on a molecular basis. Here, we examined the analgesic/adverse effect profile of a new NT(8-13) derivative denoted JMV2009, in which the $Pro^{10}$ residue was substituted by a silicon-containing unnatural amino acid silaproline. We first report the synthesis and *in vitro* characterization (receptor-binding affinity, functional activity and stability) of JMV2009. We next examined its analgesic activity in a battery of acute, tonic and chronic pain models. We finally evaluated its ability to induce adverse effects associated with chronic opioid use, such as constipation and analgesic tolerance or related to $NTS_1$ activation, like hypothermia. In *in vitro* assays, JMV2009 exhibited high binding affinity for both $NTS_1$ and $NTS_2$, improved proteolytic resistance as well as agonistic activities similar to NT, inducing sustained activation of p42/p44 MAPK and receptor internalization. Intrathecal injection of JMV2009 produced dose-dependent antinociceptive responses in the tail-flick test and almost completely abolished the nociceptive-related behaviors induced by chemical somatic and visceral noxious stimuli. Likewise, increasing doses of JMV2009 significantly reduced tactile allodynia and weight bearing deficits in nerve-injured rats. Importantly, repeated agonist treatment did not result in the development of analgesic tolerance. Furthermore, JMV2009 did not cause constipation and was ineffective in inducing hypothermia. These findings suggest that NT drugs can act as an effective opioid-free medication for the management of pain or can serve as adjuvant analgesics to reduce the opioid adverse effects.

## 1. Introduction

Today, opioids remain at the forefront of the pharmacological treatment of chronic pain (de Leon-Casasola, 2013; Gomes et al., 2014). However, opioid therapy is associated with a wide range of side effects, such as constipation, nausea, drowsiness, dizziness, sleep disturbance, dry mouth, respiratory depression and tolerance that negatively affects patients' well-being and even discourages them for continuing their medications (Benyamin et al., 2008; Bruneau et al., 2018; Devulder et al., 2005; McNicol et al., 2003). In the ongoing opioid crisis, involving drug misuse, abuse, overdose and death (Coyle et al., 2018), there is therefore an urgent need for new, effective and safe non-opioid analgesics to improve the care of patients suffering from chronic pain (Besserer-Offroy and Sarret, 2019).

Among the alternative to opioids, neurotensin (NT) receptors recently emerged as attractive targets to develop new painkillers (Perez de Vega et al., 2018). Through its binding with two G protein-coupled receptors (GPCRs), referred to as $NTS_1$ and $NTS_2$, the tridecapeptide NT(1-13) was indeed found to induce a potent opioid-independent analgesia in a variety of pain paradigms following systemic or central administration (Dobner, 2006; Feng et al., 2014; Kleczkowska and Lipkowski, 2013; Sarret and Cavelier, 2018). Accordingly, it was pharmacologically demonstrated that the antinociceptive responses induced by NT and its analogs, designed and synthesized based on the minimal biologically active C-terminal region ($Arg^8$-$Arg^9$-$Pro^{10}$-$Tyr^{11}$-$Ile^{12}$-$Leu^{13}$), could not be reversed by the opioid antagonists, naloxone and naltrexone (Behbehani and Pert, 1984; Bredeloux et al., 2006; Osbahr et al., 1981; Wustrow et al., 1995). Importantly, NT was also reported to provide stronger pain relief than morphine on a molecular basis (al-Rodhan et al., 1991; Nemeroff et al., 1979).

In the last two decades, peptide therapeutics have gained increased interest from the biomedical and pharmaceutical industry, filling the gap between the two main classes of available market drugs, small molecules and biologics (Fosgerau and Hoffmann, 2015; Kaspar and Reichert, 2013). Peptides are definitely attractive drug candidates due to their high potency/efficacy, target specificity and safety/tolerability profile. Despite offering many advantages, their extensive use as therapeutics is, however, limited by their short half-life and rapid plasma clearance, which leads to low bioavailability (Fosgerau and Hoffmann, 2015; McGonigle, 2012). Several strategies have therefore been developed to overcome the undesirable physicochemical properties of therapeutic peptides. These approaches successfully applied to the C-terminal NT(8-13) hexapeptide include backbone and side-chain modifications, such as D-amino acid and non-natural amino acid substitutions, incorporation of reduced bonds, lactam bridged mimetics or peptide-peptoid hybrids, N- and C-terminal modifications as well as introduction of conformational constraints via cyclization (Bittermann et al., 2004; Boules et al., 2010; Bredeloux et al., 2008; Einsiedel et al., 2011; Fanelli et al., 2015; Hughes et al., 2010; Labbé-Jullié et al., 1994; Sarret and Cavelier, 2018; Sousbie et al., 2018; Tyler-McMahon et al., 2000; Wustrow et al., 1995).

According to recent structure-activity relationship (SAR) studies, structural modifications with unnatural amino acids were effective in producing NT(8-13) analogs with improved metabolic stability, receptor selectivity and *in vivo* activity (Boules et al., 2010; Fanelli et al., 2015; Hapău et al., 2016; Hughes et al., 2010; Magafa et al., 2019; Tokumura et al., 1990; Tyler-McMahon et al., 2000). In that respect, non-natural silylated amino acids are of great interest for their ability to modulate the physicochemical properties of bioactive peptides (Cavelier et al., 2004; Cavelier et al., 2002; Dalkas et al., 2010; Fanelli et al., 2017; Fanelli et al., 2015; Remond et al., 2016). Herein, we report the synthesis and *in vitro/in vivo* characterization of a new NT(8-

13) analog, named JMV2009, which binds to both $NTS_1$ and $NTS_2$ receptors. In this compound, the $Pro^{10}$ residue was substituted by a silicon-containing unnatural amino acid proline surrogate, denoted silaproline (Sip), which has been shown to retain peptide conformation (Cavelier et al., 2006; Cavelier et al., 2002; Remond et al., 2017; Vivet et al., 2000). This Sip-containing NT(8-13) analog was further screened in a battery of pain models to assess its putative analgesic properties and evaluate for its ability to induce adverse effects associated with chronic opioid therapy, such as constipation or analgesic tolerance.

## 2. Materials and methods

### *2.1. Materials*

PD149163 was obtained by the U.S. National Institute of Mental Health, Chemical Synthesis and Drug Supply Program. Morphine sulfate was purchased from Sandoz (Boucherville, QC, Canada).

### *2.2. Synthesis and in vitro characterization of JMV2009*

Synthesis, chemical characterization, and *in vitro* and *in cellulo* characterization (binding, signaling and plasma stability) of JMV2009 are reported into the accompanying Data in Brief article (Besserer-Offroy et al., 2020)

### *2.3. Animals*

Experimental procedures were approved by the Animal Care and Ethical Committee of the Université de Sherbrooke, for experiments involving rats (animal use protocol number 035-18) and were in accordance with policies and directives of the Canadian Council on Animal Care. Experiments involving mice were approved by the Animal Care and Ethical Committee of the Université de Rouen (Cenomax, Agreement of the French Ministry of Research number 54) and were in accordance with policies and directives of the French Council on Animal Care and European Community Council Directive of 24 November 1986 (86:609: EEC). Furthermore, all procedures involving animals followed the ARRIVE recommendations (Kilkenny et al., 2010).

Adult male Sprague-Dawley rats (200-225 g; Charles River Laboratories, St-Constant, Quebec, Canada) or adult male Swiss albino mice (CD1, 22-26 g, Charles River, St-Aubin-les-Elbeuf, France) were both maintained on a 12:12 hrs light/dark cycle with access to food and water *ad libitum*. Animals were acclimated for four days to the animal facility and for two days to the manipulations and testing devices prior to behavioral studies. The total number of animals used in this study is 30 mice and 172 rats. Animals were independently and randomly assigned to a saline or treatment group prior any behavioral testing.

### ***2.4. Behavioral studies involving mice***

#### *2.4.1. Intracerebroventricular injections*

The intracerebroventricular injections (i.c.v.) were performed in conscious mice according to the method of (Haley and McCormick, 1957) in a volume of 10 µl/mouse.

#### *2.4.2. Visceral pain test*

The intraperitoneal (i.p.) injection of a chemical irritant in mice induces a visceral noxious response characterized by a wave of constriction followed by elongation of the abdominal wall (Collier et al., 1968). This tonic visceral nociceptive response is often described as the writhing test (Koster et al., 1959). Briefly, mice received an intracerebroventricular (i.c.v.) injection of saline or JMV2009 20 min before i.p. administration of an acetic acid solution (0.5%). The number of writhes produced for a 5-min period was then counted beginning 5 min after i.p. injection.

### ***2.5. Behavioral studies involving rats***

#### *2.5.1. Intrathecal injections*

Rats were lightly anesthetized with 2% isofluorane. Subsequently, a 25 µl volume of compound or vehicle was administered by injection into the subarachnoid space between lumbar vertebrae L5 and L6, using a 27 G 1/2 needle.

#### *2.5.2. Acute pain test (Tail-Flick test)*

Acute nociception was investigated using the tail-flick test. Thermal threshold latencies were determined at baseline and at 10, 20, 30, 40, 50, 60, and 90 min after intrathecal (i.t.) injection of JMV2009 or vehicle. Testing involved measuring the tail withdrawal latency (in s) after immersion in a 52˚C water bath. Six centimeters of the rat's tail were submerged in water and the latency to flick or curl the tail was recorded. Baseline responses (before drug administration) were typically ranging from 3 to 4 s. A cut-off was imposed at 10 s to avoid tissue damage.

The mean latency at peak effect (40 min) was used for the determination of the analgesic efficacy and was converted to % of maximal possible effect (MPE) according to the formula: % maximum possible effect = [(test latency) − (baseline latency)]/[(cut-off) − (baseline latency)] × 100.

### 2.5.3. *Formalin tonic pain test*

Persistent pain was assessed using the formalin test. For this purpose, rats were placed for a 60-min habituation period in the experimentation room two days before the test. On the test day, rats received 50 µl subcutaneous injection of diluted 2% formaldehyde into the plantar surface of the right hind paw five min. after i.t. injection of JMV2009 or saline. Rats were then immediately placed in clear plastic chambers (40 × 30 × 30 cm) positioned over a mirror angled at 45°, to allow an unobstructed view of the 4 paws. Behaviors were recorded for the next 60 min.

An intra-plantar injection of formalin produces a biphasic nociceptive response that characterizes this tonic pain model (Sawynok and Liu, 2004). The two distinct phases of spontaneous pain behaviors that occur in rats are suggested to reflect a direct effect of formalin on sensory receptors (phase 1) and a longer lasting pain due to inflammation and central sensitization (phase 2). The nocifensive behaviors were assessed using a weighed score, as described previously by (Dubuisson and Dennis, 1977). A nociceptive mean score was determined for each 3-min period by measuring the amount of time spent in each of four behavioral categories: 0, the injected paw is comparable to the contralateral paw; 1, the injected paw has little or no weight placed on it; 2, the injected paw is elevated and is not in contact with any surface; 3, the injected paw is licked, bitten, or shaken. The total AUC for the inflammatory phase (phase 2) was calculated between 21 to 60 min for each animal (Gaumond et al., 2007).

### *2.5.4. Chronic Construction Injury (CCI) of the sciatic nerve*

#### *2.5.4.1. Surgical procedure*

The surgical procedure of the neuropathic pain model was performed as previously described by (Bennett and Xie, 1988) with a modification in the suture used (5-0 Prolene, Ethicon, Inc. Somerville, NJ, USA) in order to minimize suture-induced inflammation. Briefly, animals were anesthetized using 3% isoflurane and the left sciatic nerve was loosely ligatured with 4 sutures distant by 1 mm upstream of the tibial, sural, and common peroneal nerve trifurcation. The muscle, conjunctive tissue and skin were closed with proper sutures and washed with 3% (v/v) hydrogen peroxide. Sham animals received the surgery, but no ligation of the sciatic nerve was performed. Rats were housed for 24 h in separate cages to recover from the surgery and were not given any medication to control post-surgical pain.

The pain-related behaviors were examined at day 0 (pre-surgery, considered the baseline [BL]) and at 3, 7, 14, 21, and 28 days after nerve injury. Animals were moved to the examination room at 8 a.m. and tested one h later. Dynamic weight bearing (DWB) measurements were conducted immediately after the mechanical allodynia assessment.

#### *2.5.4.2. Evaluation of mechanical allodynia*

To determine the presence of mechanical allodynia, rats were placed in enclosures with an elevated wire mesh floor. A dynamic plantar aesthesiometer (Ugo Basile, Stoelting, IL, USA), consisting of a metal probe (0.5 mm diameter), was directed against the hind paw pad exerting an upward force (3.33 g/s). The force required to elicit a withdrawal response was measured in grams and automatically registered when the paw was withdrawn, or the pre-set cut-off was reached (50 g). Five values were taken alternately on ipsi- and contra-lateral hind paw at 15 s intervals. The percentage of anti-allodynia was calculated with the AUC of every treatment for the time period comprised between 3 and 28 days, with the use of the following equation: % anti-allodynia =100 × [(CCI + drug) − CCI control]/(sham − CCI control). From the latter formula, 0% represents no anti-allodynic effect of the compound, while 100% corresponds to a complete relief of mechanical hypersensitivity.

*2.5.4.3. Dynamic weight bearing measurements*

The DWB device (Bioseb, Boulogne, France) was used to evaluate the presence of non-evoked pain. This apparatus was characterized previously by our group (Tétreault et al., 2011). Briefly, rats are allowed to move freely within a Plexiglas enclosure (22 × 22 × 30 cm) with a floor sensor composed of 44 × 44 captors (10.89 $mm^2$ per captor). Pressure data and live recording are recorded and transmitted to a laptop computer. Raw pressure and visual data were colligated with the latest DWB software available at the time. To calculate weight recovery and rehabilitation, we used the same formula as the anti-allodynia, meaning that: % weight recovery (or rehabilitation) = 100 × [AUC(CCI + drug) – AUC(CCI control)]/[(AUC(sham) – AUC(CCI control)]. From this formula, 0% represents no weight recovery (or rehabilitation) induced by the compound, while 100% corresponds to complete weight recovery or total rehabilitation.

## *2.6. Monitoring of the adverse effects*

### *2.6.1. Gastro-intestinal motility assessment*

Constipation has been assessed by measuring the gastro-intestinal (GI) tract motility using the charcoal meal test. Food deprived (16 h) animals were injected s.c. with saline or the compounds to be tested. 30 min. after drug injection, 2-ml of a charcoal meal solution (5% arabic gum, 10% charcoal in water) was administered to the rats by gavage. Rats were euthanized exactly 60 min after, and the progression of the charcoal in the intestine was measured as a ratio of progression of the meal over the total length of the intestine. The results are presented as a percentage of progression of the charcoal meal in the intestine.

*2.6.2. Tolerance to the acute thermal pain test*

For tolerance experiments, rats were injected i.t. with 30 µg/kg/day of JMV2009 or with saline for five consecutive days. On the sixth day, tail-flick latencies were measured immediately before and over 90 min following i.t. injection of JMV2009 (90 µg/kg). The analgesic effect of this sub-chronic treatment was assessed using the tail-flick thermal sensitivity assay as described above.

*2.6.3. Blood pressure monitoring*

Male Sprague-Dawley rats were anesthetized with a mixture of ketamine/xylazine (87 mg/kg:13 mg/kg, i.m.) and placed in supine position on a heating pad. Mean, systolic and diastolic arterial blood pressure as well a heart rate were measured through a catheter (PE 50 filled with heparinized saline) inserted in the right carotid artery and connected to a Micro-Med transducer (model TDX-300, USA) linked to a blood pressure Micro-Med analyzer (model BPA-100c). Another catheter was inserted in the left jugular vein for bolus injections (1 ml/kg, 5-10 s) of vehicle (isotonic saline), NT[8-13] or JMV2009. For relative potency evaluation, changes in mean arterial blood pressure (ΔMABP) from baseline to post-injection in individual animals were determined.

*2.6.4. Core body temperature*

Body temperature was measured using a thermistor probe inserted into the rectum of adult Sprague-Dawley rats. Prior to testing, animals were individually acclimatized to manipulations and thermistor probe 5 min/day for three consecutive days. On the test day, temperature was measured before (baseline) and each 30 min for up to 90 min following i.t. injection of saline, JMV2009 (3, 30 and 90 µg/kg) or the $NTS_1$-selective reference compound PD149163 (Feifel et al., 2010). Variations in body temperature (Δ body temp) were determined as changes from baseline for each individual animal.

### *2.7. Data analysis and statistical procedures*

Data were plotted as mean ± standard error (S.E.M.) for all curves and bar graphs. Two-way ANOVAs followed by a Dunnett's post hoc test were performed for all curves. Kruskal-Wallis followed by a Dunn’s correction for multiple comparisons tests were performed for all bar graphs except for the weight recovery and rehabilitation where a Mann-Whitney rank comparison test was used. Prism v7.0a software was used for all statistical analyses. A $P$ value < 0.05 was considered significant and significant differences between groups were represented by * $P < 0.05$, ** $P < 0.01$, and *** $P < 0.001$. Dose-response curve was analyzed by using the log(agonist) vs. response (three parameters) non-linear regression fit of GraphPad Prism v7.0a.

## 3. Results

### *3.1. JMV2009 acts as a non-selective NT receptor agonist*

We report here the synthesis and pharmacological evaluation of a NT(8-13) analog (JMV2009) carrying an unnatural silicon-containing proline surrogate, the 4-(dimethyl)silaproline (Sip) in replacement of the proline residue at position 10 (***Fig. 1-2 and Scheme 1 of the accompanying Data in Brief article***) . We first found that this silaproline-containing peptide exhibited relatively high and similar binding affinities for both $NTS_1$ and $NTS_2$ receptors, with only slightly reduced affinity compared to the native NT peptide (***Fig. 3 and Table 1 of the accompanying Data in Brief article***). Our results further demonstrated that the presence of this silylated proline residue at position 10 can increase peptide resistance to proteolytic degradation. Accordingly, around 80 % of JMV2009 remained intact after a 5-min incubation with rat plasma, compared to 15 % for the native NT peptide, thus resulting in an improved peptide stability to proteases (***Fig. 4 of the accompanying Data in Brief article***).

We next examined if this NT peptide derivative was still able to induce receptor activation, by studying its ability to stimulate known intracellular signaling pathways activated by the NT receptor family. We first evaluated the ability of JMV2009 to trigger the activation of different G proteins ($G\alpha_q$, $G\alpha_{13}$, $G\alpha_{i1}$, and $G\alpha_{oA}$) and the recruitment of β-arrestins 1 and 2 as it was previously reported for NTS1 endogenous ligands (Besserer-Offroy et al., 2017). We found that JMV2009 was able to activate the four studied G proteins as well as to promote the recruitment of both β-arrestins 1 and 2 (***Fig. 5 and Table 2 of the accompanying Data in Brief article***). We next evaluated the phosphorylation level of the mitogen-activated protein kinase ERK1/2 after NT or JMV2009 stimulation using cell lines stably expressing either $NTS_1$ or $NTS_2$ receptors. Our results revealed that JMV2009 was able to induce ERK1/2 activation in a similar manner as NT in both $NTS_1$- and $NTS_2$-expressing cells (***Fig. 6 of the accompanying Data in Brief article***). We then assessed the ability of JMV2009 to trigger NT receptor internalization by cell surface enzyme-linked immunosorbent assay (***Fig. 7 and Table 3 of the accompanying Data in Brief article***). We observed that both NT and JMV2009 promoted $NTS_1$ receptor internalization, resulting respectively in 59.4 ± 1.5% and 49.1 ± 1.4 % reduction in receptor cell surface expression within 1 h. Likewise, NT and JMV2009 induced $NTS_2$-receptor internalization after 60 min of stimulation. Altogether, these results demonstrate that the substitution of $Pro^{10}$ by Sip leads to a functional NT agonist that might be associated with antinociceptive properties.

### *3.2. Central delivery of JMV2009 significantly reduces acute and tonic pain*

We first used the tail-flick acute pain model to evaluate whether intrathecal (i.t.) delivery of JMV2009 attenuated the withdrawal responses to thermal nociceptive stimuli. We found that i.t. delivery of JMV2009 in rats elicited a dose-dependent antinociceptive response up to 60 min after injection, characterized by an increased tail-flick latency compared to saline-treated animals (**Fig. 1A**). Peak analgesic responses occurred 40 min after JMV2009 injection, tail-flick latencies returning to baseline by 90 min. Comparison of % MPE at 40 min following JMV2009 administration showed significant antinociceptive effects for each dose tested. Antinociception reached 35.2 ± 3.2 %, 43.4 ± 8.9 %, and 62.0 ±7.7 % at doses of 3, 30 and 90 µg/kg, respectively (**Fig. 1B**).

We then assessed the efficacy of JMV2009 to alleviate the nociceptive behaviors in two different persistent pain paradigms, using either intraplantar formalin or intraperitoneal injection of acetic acid as chemical somatic and visceral noxious stimuli, respectively. Intraplantar injection of formalin into the right hind paw of saline-pretreated rats induced a biphasic time-dependent increase in pain score (**Fig. 2**). Importantly, i.t. pretreatment with JMV2009 significantly reduced the stereotypical nocifensive behaviors elicited by formalin during the tonic inflammatory phase (21 to 60 min), without affecting the early phase (1 to 9 min) (**Fig. 2A**). Indeed, both doses tested markedly abolished the spontaneous persistent pain-related behaviors, reaching 49.7 % and 97 % of inhibition at 3 and 30 µg/kg, respectively (**Fig. 2B**). The antinociceptive effect of JMV2009 was further tested using the acetic acid-induced writhing model in mice (**Fig. 2C**). We found that i.c.v. injection of JMV2009 suppressed the visceral pain behaviors in a dose-dependent manner, reducing the number of writhes by 87 % at the highest dose tested and displaying a calculated $ED_{50}$ value of 2.33 ± 0.6 µg/kg (**Fig. 2D**).

### *3.3. JMV2009 attenuates neuropathy-induced mechanical hypersensitivity and improves the quality-of-life proxies*

We next investigated the antinociceptive action of JMV2009 in a clinically relevant model of chronic neuropathic pain. Neuropathic pain was induced by chronic constriction injury (CCI) of the sciatic nerve and the mechanical hypersensitivity was determined by applying von frey filaments. As shown in **Fig. 3**, CCI rats developed tactile allodynia as soon as 3 days after surgery compared with preoperative values and persisted until at least 28 days. As expected, the paw withdrawal thresholds in response to tactile stimuli were not affected in sham-operated animals. Importantly, i.t. injection of JMV2009 was effective in reversing the allodynic state, even at the lowest dose tested (**Fig. 3A**). Indeed, JMV2009 produced robust anti-allodynic effects, achieving 51.4 ± 3.4 %, 69.2 ± 3.9 % and 66.2 ± 7.6 % of pain relief at the doses of 3, 30 and 90 µg/kg, respectively (**Fig. 3B**).

To date, the development of potential analgesic agents relies mainly on the measurement of withdrawal responses to acute application of external heat or mechanical stimuli (e.g., von frey filament). However, to improve the translation of analgesics to the clinic, it may be useful to monitor spontaneous innate behaviors indicative of animal well-being in the preclinical drug screening. Accordingly, we previously demonstrated in different chronic pain models that the dynamic weight bearing (DWB) method could be used to measure posture-locomotor functional impairment and rehabilitation as well as health and quality-of-life outcomes in freely moving rats (Tetreault et al., 2013; Tétreault et al., 2011). In addition to the reflexive mechanical nociceptive test, we thus used the DWB device to assess the effects of JMV2009 on weight load deficits and time-of-use of the injured paw. We first found that i.t. delivery of JMV2009 was effective to reverse the weight bearing deficit induced by the chronic constriction injury of the sciatic nerve (**Fig. 4A**). Indeed, a 30-µg/kg dose of JMV2009 significantly reduced the ipsilateral paw weight load deficits in non-restrained neuropathic rats, leading to a gain of 40.9 ± 6.4 % of weight recovery over the 28-day period of observation (**Fig. 4A-B**). Next, we evaluated the efficacy of JMV2009 to reverse the impaired use of the affected limb. As shown in **Fig. 4C**, JMV2009 improved the time spent by the animal on its injured paw. The drug effectiveness was also expressed as % of rehabilitation within the four-week period, where 100% rehabilitation corresponds to a complete recovery of the paw usage. This outcome parameter is considered clinically relevant and represents a good indicator of quality of life improvement (Cobos et al., 2012). At the dose used, JMV2009 was found to improve the use of the affected limb to 52.8 ± 10.3% (**Fig. 4D**).

### *3.4. JMV2009 displays a safer profile compared to opioids or other NT derivatives*

To achieve good analgesia, patients' compliance to a prescribed pain treatment is of high importance. Among the various reasons of noncompliance, the presence of mild to severe side effects seriously affects medication adherence in patients with chronic pain (Benyamin et al., 2008). For instance, constipation, nausea, respiratory depression and tolerance are common clinical concerns that may lead to discontinuation of long-term opioid therapy. Here, we therefore assessed whether JMV2009 could induce constipation after systemic administration, as it is reported that NT and NT derivatives could have a direct effect on smooth muscles of the intestine in isolated ileum organ bath (Fanelli et al., 2015). To this end, we measured the effects of JMV2009 and morphine on the gastrointestinal tract motility using the charcoal meal test in starved rats (**Fig. 5A**). Morphine (10 mg $kg^{-1}$ s.c.) was found to reduce by 45% the progression of the charcoal meal in the intestine, when compared to saline-treated animals. By contrast, both NT and JMV2009 subcutaneously injected at 10 mg $kg^{-1}$ did not affect the charcoal meal intestinal transit. Rats treated with JMV2009 even showed an almost significant increase of the gastrointestinal tract motility, since the charcoal meal travelled 98.7 ± 0.8 % of the intestine ($P$ = 0.064) compared to 82.2 ± 2.4 % and 87.4 ± 2.5 % when rats were treated with either NT or saline, respectively (**Fig. 5A**).

We next determined whether repeated injection of JMV2009 induced the development of tolerance to its analgesic effect. For this purpose, rats were subjected to i.t. injections of JMV2009 (30 µg/kg) daily for five consecutive days, and tail-flick tests were conducted over 90 min on the last day to evaluate changes of JMV2009-induced antinociception. In this experimental paradigm, repeated JMV2009 exposure did not produce tolerance, as demonstrated by the absence of changes in tail-flick latency, when compared to saline pre-treated animals (**Fig. 5B**). Next, since NT administered centrally also causes hypothermia through an $NTS_1$-dependent mechanism (Bissette et al., 1976; Feifel et al., 2010), we measured the ability of JMV2009 to induce changes in body temperature after central administration. We found here that JMV2009 administered i.t. at effective analgesic doses did not induce hypothermia. In sharp contrast, the $NTS_1$-selective agonist PD149163, used here as a reference compound induced a pronounced time-dependent drop in body temperature (**Fig. 5C**). Finally, we investigated the ability of JMV2009 to induce changes in blood pressure after systemic injection. We found here that a dose of 0.1 mg $kg^{-1}$ of JMV2009 injected i.v. was able to induce a drop of blood pressure similar to the one induced by the same dose of NT. Changes in mean arterial blood pressure (MABP) were found to be of -29.2 ± 5.7 mmHg for rats receiving JMV2009 compared to -25.0 ± 2.5 mmHg for NT-injected animals (**Fig. 5D**).

## 4. Discussion

G-protein coupled receptors are the target of more than 25% of the FDA-approved drugs (Overington et al., 2006) and represent one of the largest classes of therapeutic targets in pain medicine (Stone and Molliver, 2009). Nowadays, validation of new molecular targets and development of new effective, non-addicting pain medications are critical, but still remain extremely slow (Kissin, 2010; Yaksh et al., 2018). In the past decades, both $NTS_1$ and $NTS_2$ have generated a growing interest as potential targets for pain treatment (Dobner, 2006; Feng et al., 2014; Kleczkowska and Lipkowski, 2013; Sarret and Cavelier, 2018). This increasing attention is probably driven in part by the fact that NT drugs act as effective opioid-free medications for pain management, producing significant pain relief in various animal models of acute and chronic pain, including neuropathic pain for which opioids have shown their limitations (Guillemette et al., 2012; Perez de Vega et al., 2018; Tetreault et al., 2013). Importantly, the combination of NT analogs to opioids has also the potential to exert complementary (synergistic or additive) analgesic actions, thereby minimizing the adverse effects related to chronic opioid use (Boules et al., 2011; Boules et al., 2009; Eiselt et al., 2019).

In recent years, changes in the pharmaceutical industry have led to renewed interest in the use of peptides as therapeutics (Fosgerau and Hoffmann, 2015; Kaspar and Reichert, 2013). The current pipeline of peptide therapeutics includes over 60 approved drugs and 150 in active clinical development, with 40% of these peptide drug candidates targeting GPCRs (Lau and Dunn, 2018). The development of peptide drugs with high therapeutic potency remains however challenging and requires the implementation of chemical strategies and/or computational structural prediction to synthesize modified peptides with improved stability, pharmacokinetics and *in vivo* activity profiles. Among the chemical approaches used to optimize the properties of the NT(8-13) peptide, incorporation of unnatural amino acids, such as N-methylarginine, D-ornithine, D-$\alpha$-naphthylalanine, *tert*-leucine or L-(trimethylsilyl)alanine respectively at positions 8, 9, 11, 12 and 13, has been shown to be effective, increasing significantly $NTS_1$ and $NTS_2$ receptor binding affinity or selectivity as well as peptide stability and *in vivo* biological activity (Boules et al., 2010; Fanelli et al., 2015; Hughes et al., 2010; Rossi et al., 2010; Tokumura et al., 1990; Tyler-McMahon et al., 2000). To date, only few studies have investigated the impact of $Pro^{10}$ substitutions on NT receptor binding and to our knowledge, none of them have evaluated their efficacy *in vivo*. The proline residue in position 10 plays a crucial role for peptide conformation and SAR studies reveal that substituted proline analogs that promote a reverse turn are more tolerated than those inducing an extended backbone conformation (Bittermann et al., 2004; Held et al., 2013; Heyl et al., 1994). Indeed, replacement of $Pro^{10}$ by thioproline, hydroxyproline, its 4 and 6 member ring counterparts azetidine carboxylic acid and pipecolic acid, or Tic and Aic cyclic aromatic derivatives all induces conformational changes resulting in a significant loss of NT receptor binding affinity (Bittermann et al., 2004; Heyl et al., 1994). These results are consistent with the crystal structure of the $NTS_1$-NT(8-13) complex in its activated-like conformation, which shows a tight oriented hydrophobic binding site for the pyrrolidine ring of $Pro^{10}$ as well as extensive van der Waals interactions with the residue W339 in ECL3 (Krumm et al., 2016; White et al., 2012). Here, we have therefore replaced $Pro^{10}$ by a silylated proline surrogate that exhibits very similar conformational properties that the proline residue in peptides (Remond et al., 2016, 2017). The substitution of the γ-carbon by a dimethylsilyl group was found to slightly decrease the affinity of JMV2009 for $NTS_1$ while this chemical substitution seemed to be adequate for $NTS_2$ binding. Accordingly, proline isostere bearing a thiazolidine unit instead of the pyrrolidine ring, as well as fluoro-substituted proline analogs on $C_\gamma$ have already been reported to be well tolerated when incorporated into $NTS_2$-selective peptoids (Held et al., 2013).

This Sip-containing NT(8-13) peptide also exhibited an improved resistance to proteolytic degradation, probably to the endopeptidases 24.11 and 24.16 responsible for inactivating NT by cleaving at the $Pro^{10}$-$Tyr^{11}$ bond (Labbé-Jullié et al., 1994; Myers et al., 2009). The steric hindrance of the dimethyl group borne by the silicon atom could indeed induce a misrecognition of the $Sip^{10}$-$Tyr^{11}$ peptide bond by these metallo-endopeptidases. To further increase the plasma stability of this NT derivative, it would be interesting to combine the substitution of $Pro^{10}$ by Sip to N-terminal protection through the incorporation of blocked amino group (i.e. $N_{\alpha}$-methylarginine) at position 8 or reduced amide bond between positions 8 and 9 as well as to replace $Ile^{12}$ by a *tert*-leucine residue to avoid enzymatic cleavage at the $Tyr^{11}$-$Ile^{12}$ peptide bond (Dubuc et al., 1992; Fanelli et al., 2015; Tokumura et al., 1990). Importantly, this Sip-containing NT(8-13) analog was effective in triggering intracellular signaling responses at both $NTS_1$ and $NTS_2$, as shown by the activation of G proteins, recruitment of β-arrestins, ERK1/2 activation and receptor internalization (Besserer-Offroy et al., 2017; Besserer-Offroy et al., 2020; Gendron et al., 2004).

We then reported the analgesic efficacy of this silylated NT(8-13) analog in different experimental pain models. We first found that JMV2009 was able to significantly reduce acute thermal pain, even at the lower dose of 3 μg/kg. This result is consistent with previous studies demonstrating that both $NTS_1$ and $NTS_2$ agonists decrease the tail and foot withdrawal nociceptive responses to noxious heat stimuli (Boules et al., 2009; Hughes et al., 2010; Rossi et al., 2010; Sarret et al., 2005). In persistent pain paradigms, JMV2009 was also able to totally abolish the nociceptive behaviors induced by either intraplantar injection of formalin or visceral noxious stimulation. Interestingly, JMV2009 displays more potent analgesic effects in the formalin pain model than the other non-selective NT analogs, NT69L and JMV2007 previously described (Fanelli et al., 2015; Roussy et al., 2008). At equimolar dose, it is also more potent than the metabolically stable $NTS_2$-selective analog JMV431 in reducing the formalin-induced nocifensive behaviors (Roussy et al., 2009). In the peripheral neuropathic pain model, JMV2009 induced relief of CCI-induced evoked hypersensitivity and ongoing pain. Indeed, we first observed that acute i.t. administration of JMV2009 was effective in reversing the development of tactile allodynia in nerve-injured rats. These data thus reinforce previous findings demonstrating that central or systemic delivery of either $NTS_1$ or $NTS_2$ agonists reduces the mechanical hypersensitivity in preclinical models of neuropathic pain (Demeule et al., 2014; Guillemette et al., 2012; Hughes et al., 2010; Tetreault et al., 2013). Finally, and perhaps most importantly, the spinal delivery of JMV2009 was also able to reverse significantly the weight bearing deficits as well as to improve the rehabilitation outcomes. In freely moving rats, we indeed found that JMV2009 induced partial recovery of the weight load on the injury leg and increased the time-of-use of the affected limb. The monitoring of these spontaneous nociceptive-related behaviors is now considered critical for evaluating the effectiveness of new drugs (Deuis et al., 2017; Mogil, 2009; Tappe-Theodor and Kuner, 2014). Most behavioral tests currently used rely only on the measurement of stimuli-evoked pain behaviors, such as mechanical hyperalgesia and allodynia. However, neuropathic pain patients predominantly suffer from non-evoked ongoing and spontaneous pain, and not from stimulus-evoked hypersensitivity. This type of pain is the one having the most negative impact, seriously affecting the patient's daily activities and quality of life, and leading to multiple comorbid conditions, such as mood and anxiety disorders, sleep perturbation, depression or attention deficits. Compared to the previously reported NT(8-13) analog, JMV431, JMV2009 produces stronger anti-allodynic effects in CCI-treated animals as well as substantial improvement in the management of the health and quality-of-life outcomes (Tetreault et al., 2013). Finally, when compared to typical first- and second-line medications for neuropathic pain, JMV2009 appeared as potent as the tricyclic antidepressant amitriptyline and more effective than morphine, pregabalin, gabapentin and ibuprofen in relieving spontaneous pain behaviors (Angeby-Moller et al., 2008; Deseure and Hans, 2017; Nakazato-Imasato and Kurebayashi, 2009; Tetreault et al., 2013).

Adherence to long-term opioid therapies represents a very clinical issue. Indeed, although opioids, such as morphine have been shown to be effective in relieving pain, the development of bothersome side effects may outweigh the benefits of opioid treatment and even lead to prompt discontinuation of the drug treatment (Benyamin et al., 2008; Besserer-Offroy and Sarret, 2019). Constipation is particularly difficult to manage since patients don't usually develop tolerance upon repeated morphine administration (Bowers and Crannage, 2017). Likewise, tolerance to the respiratory depressant effects of opioids develops only slowly and incompletely, putting patients at risk for respiratory depression with dose escalation (Algera et al., 2019). Furthermore, the long-term use of opioids can lead to analgesic tolerance and opioid-induced hyperalgesia (Chu et al., 2006). Unlike morphine, we found here that both NT and JMV2009 did not cause constipation. On the contrary, JMV2009 seemed to induce better gastrointestinal transit compared to NT-injected rats. These results are consistent with previous studies demonstrating that NT stimulates colonic motility in rats, cats and humans (Hellstrom and Rosell, 1981; Pellissier et al., 1996; Thor and Rosell, 1986). Supporting a role for NT in propulsive colonic motility is also the finding that i.v. injection of NT in human subjects stimulates defecation and may improve the intestinal movements in colons of patients with slow transit constipation (Calam et al., 1983; Tomita et al., 2007). We further demonstrated that JMV2009 did not produce analgesic tolerance following repeated agonist administration. Accordingly, various NT analogs, such as [D-Trp$^{11}$]-NT, ABS212 and β-lactotensin do not induce the development of antinociceptive tolerance (Dubuc et al., 1994; Hughes et al., 2010; Yamauchi et al., 2003). We also showed that JMV2009 was ineffective for inducing hypothermia, as opposed to the NT(8-13) agonist PD149163 that triggered a sustained decrease in body temperature at equimolar dose. This result is quite surprising since most NT agonists acting at $NTS_1$ can produce a significant drop in body temperature (Bissette et al., 1976; Fantegrossi et al., 2005; Feifel et al., 2010). Nevertheless, few NT analogs acting at $NTS_1$ were previously reported to be able to distinguish between hypothermia and analgesia, such as the NT(8-13) analogs, NT27 and NT77L (Boules et al., 2001; Tyler et al., 1998). The concept of ligand biased signaling provides support to the idea that NT analogs acting at $NTS_1$ may have the ability to part between analgesia and hypothermia by recruiting different receptor downstream signaling pathways (Kenakin, 2019). Finally, we showed that JMV2009, when injected systemically, was able to induce a drop in blood pressure which was comparable to the drop induced by NT. This effect, mediated by $NTS_1$, can be triggered after an i.v. injection of a non-selective neurotensinergic compound but also after i.t. delivery at higher doses (Zogovic and Pilowsky, 2012). Indeed, as shown by Zogovic and Pilowsky, administration of at least 75 µM of NT is required to lower blood pressure after spinal delivery (Zogovic and Pilowsky, 2012). For comparison purposes, the analgesic dose of 30 µg/kg of JMV2009 corresponds to an injection of less than 500 nM (for a rat weighting 300 g). Finally, NT- and opioid-related adverse effects could be mitigated by the use of combining neurotensinergic and opioidergic ligands either bi-topic ligands or combination therapy, as recently demonstrated using systemic injections of morphine combined to a brain-penetrant NT agonist (Eiselt et al., 2019; Kleczkowska et al., 2010).

## 5. Conclusion

In conclusion, we demonstrate here for the first time that the NT(8-13) peptide can accommodate the substitution of the proline residue to give rise to a powerful analgesic peptide. We further provide evidence that NT agonists do not induce the common adverse effects associated with the use of opioids (i.e. constipation and tolerance), thus offering a non-opioid option for the treatment of pain. Finally, these findings suggest that NT agonists used combined with opioid therapy might contribute to adequate pain relief, while reducing the severity of opioid- and neurotensin-driven adverse events.

**Authors contributions**

**Pascal Tétreault:** Conceptualization, Methodology, Validation, Formal analysis, Investigation, Writing – Original Draft, Writing – Review & Editing, Visualization.

**Élie Besserer-Offroy:** Conceptualization, Methodology, Validation, Formal Analysis, Investigation, Writing – Original Draft, Writing – Review & Editing, Visualization.

**Rebecca L Brouillette:** Investigation.

**Adeline René:** Ressources.

**Alexandre Murza:** Investigation.

**Roberto Fanelli:** Ressources.

**Karyn Kirby:** Investigation.

**Alexandre J Parent:** Investigation.

**Isabelle Dubuc:** Investigation.

**Nicolas Beaudet:** Conceptualization, Validation, Formal Analysis.

**Jérôme Côté:** Investigation.

**Jean-Michel Longpré:** Conceptualization, Validation, Formal Analysis.

**Jean Martinez:** Ressources, Supervision.

**Florine Cavelier:** Ressources, Writing – Review & Editing, Supervision, Funding Acquisition.

**Philippe Sarret:** Conceptualization, Validation, Formal Analysis, Supervision, Funding Acquisition.

All the authors approved the version of the manuscript to be published.

**Acknowledgements**

PT is the recipient of a post-doctoral fellowship from the Canadian Institutes for Health Research (CIHR, grant MFE-135472). ÉBO is the recipient of a Fond de recherche du Québec – Santé (FRQ-S, grant 255989) and a CIHR (grant MFE-164740) research fellowships. RLB has been awarded an FRQ-S and a Faculty of Medicine and Health Sciences of the Université de Sherbrooke Ph.D. scholarships. PS is the recipient of a Tier 1 Canada Research Chair in Neurophysiopharmacology of Chronic Pain and a member of the FRQ-S-funded Québec Pain Research Network (QPRN). The National Institute of Mental Health (NIMH) is acknowledged for providing us with the $NTS_1$-selective compound PD149163.

**Funding sources**

This research was supported by grants from the Canadian Institutes for Health Research [grant numbers MOP-74618, FDN-148413]

**Conflicts of interest**

The authors declare that they have no known competing financial interests or personal relationships which have, or could be perceived to have, influenced the work reported in this article.

## Figure legends

**Fig. 1. Antinociceptive response to i.t. injection of JMV2009 in the tail-flick test.** (**A**) Dose- and time-dependent antinociceptive effect of JMV2009 (3, 30, and 90 µg/kg) on tail-flick latencies in rats. Baseline latencies were taken three times before acute i.t. injection of JMV2009. Latencies were determined every 10 min for up to 60 min and a final measure was taken at 90 min following drug administration. ** $P < 0.01$ (JMV2009, 3 µg/kg vs. Saline); ## $P < 0.01$ (JMV2009, 30 µg/kg vs. Saline); † $P < 0.05$, and ††† $P < 0.001$ (JMV2009, 90 µg/kg vs. Saline) in a two-way ANOVA ($F(7, 393) = 25.40$) followed by a Dunnett's post hoc test. (**B**) Percentage of maximum possible effect (MPE) ± S.E.M. determined at the time of peak anti-nociceptive response (40 min) for each dose (3, 30, and 90 µg/kg). * $P < 0.05$ and *** $P < 0.001$ (JMV2009 vs. Saline) in a Kruskal-Wallis test ($\chi^2(3) = 19.48$, $P = 0.0002$) followed by a Dunn's correction for multiple comparisons. Each data point represents the mean ± S.E.M. n = 8 rats per group.

**Fig. 2. Effects of central administration of JMV2009 on tonic and visceral pain.** (**A**) Reduction of the nocifensive behaviors after i.t. injection of JMV2009 (3 and 30 μg/kg) in the formalin-induced tonic pain model. * $P < 0.05$ (JMV2009, 3 μg/kg vs. Saline) and $^{\#\#\#}$ $P < 0.01$ (JMV2009, 30 μg/kg vs. Saline) in a two-way ANOVA (F(2, 455) = 168.7) followed by a Dunnett's post hoc test. (**B**) The cumulative nociceptive response, expressed as mean area under the curve (AUC) ± S.E.M., is measured during the inflammatory phase of the formalin test (21 – 60 min). *** $P < 0.001$ (JMV2009 vs. Saline) in a Kruskal-Wallis test ($\chi^2(2)$= 17.82, $P = 0.0001$) followed by a Dunn's correction for multiple comparisons. Each symbol represents the mean ± S.E.M. of determinations made in 6 rats. (**C**) Dose-dependent antinociceptive effects of JMV2009 (0.4, 4, 12 and 40 μg/kg, i.c.v.) in the acetic acid-induced writhing model in mice. Writhes were counted over a 5-min period after i.p. injection of 0.5% acetic acid. ** $P < 0.01$ and *** $P < 0.001$ (vs. Saline), $^{\#\#}$ $P < 0.01$ ( JMV2009 40 vs. JMV2009, 0.4 μg/kg) in a Kruskal-Wallis test ($\chi^2(4)$= 31.06, $P < 0.0001$) followed by a Dunn's correction for multiple comparisons. The number of indicated writhes is the mean ± S.E.M. from groups of 6 mice. (**D**) Dose-response curve of JMV2009-induced antinociception in the acetic acid in the writhing model in mice. JMV2009 displayed an $ED_{50}$ of 2.33 ± 0.6 μg/kg, as determined by non-linear regression fit in GraphPad Prism ($R^2 = 0.61$).

**Fig. 3. Effects of JMV2009 on mechanical allodynia induced by the chronic constriction injury (CCI) of the sciatic nerve in rats.** (**A**) Paw withdrawal thresholds were assessed with an automated von Frey hair during the 4-week period after CCI surgery. Baseline (BL) withdrawal thresholds were determined for all CCI-and sham-operated rats prior to surgery. In neuropathic rats, acute i.t. injection of JMV2009 (3, 30, and 90 µg/kg) at day 3, 7, 14, 21, and 28 post-surgery effectively reduces the mechanical hypersensitivity. ** $P < 0.01$, *** $P < 0.001$ (JMV2009, 3 µg/kg vs. CCI); $^{###}$ $P < 0.001$ (JMV2009, 30 µg/kg vs. CCI); $^{†††}$ $P < 0.001$ (JMV2009, 90 µg/kg vs. CCI) in a two-way ANOVA (F(4, 182) = 236.8) followed by a Dunnett's post hoc test. (**B**) Percentage of anti-allodynia calculated over the 3-28 day period. All tested doses of JMV2009 were effective to attenuate the development of mechanical allodynia. *** $P < 0.001$ (JMV2009 vs. Saline) in a Kruskal-Wallis test ($\chi^2(3)$= 26.59, $P < 0.0001$) followed by a Dunn's correction for multiple comparisons. Vertical bars represent S.E.M. (n = 6 per group).

**Fig. 4. Effects of JMV2009 on neuropathy-induced weight-bearing deficits and decrease in time-of-use of the injured limb.** (**A**) Weight bearing impairment was measured using the dynamic weight bearing (DWB) after acute i.t. injection of JMV2009 (30 µg/kg) at day 3, 7, 14, 21, and 28 post-surgery. Baseline values (BL) were assessed before CCI surgery. * $P < 0.05$ and ** $P < 0.01$ (JMV2009, 30 µg/kg vs. CCI) in a two-way ANOVA ($F(2, 30) = 61.79$) followed by a Dunnett's post hoc test. (**B**) Treatment effectiveness was determined by measuring the percentage of weight recovery over the 3-28 day period, with 100% corresponding to full recovery in the weight borne on the injured limb. *** $P < 0.001$ in a two-tailed Mann-Whitney test. (**C**) Percentage of time spent on the injured limb was determined using the DWB device over the 28-day period following CCI surgery. * $P < 0.05$ (JMV2009, 30 µg/kg vs. CCI) in a two-way ANOVA ($F(2, 30) = 19.19$) followed by a Dunnett's post hoc test. (**D**) Determination of the percentage of functional rehabilitation was examined over the 28-day period after acute administration of JMV2009. 100% rehabilitation corresponds to a full recovery in the use of the injured limb. * $P < 0.05$ in a two-tailed Mann-Whitney test. Each symbol represents the mean ± S.E.M. of determinations made in 6 rats.

**Fig. 5. Effects of JMV2009 on opioid and neurotensin-related adverse effects.** (**A**) Gastrointestinal tract motility assessed by measuring the progression of a charcoal meal in the intestine. Results are expressed as percentage of distance travelled by the active charcoal 30 min after the s.c. injection of either saline (1 ml/kg), Morphine sulfate, neurotensin (NT), or JMV2009 (10 mg/kg). *** $P < 0.001$ (vs. Saline) in a Kruskal-Wallis test ($\chi^2(3)= 18.95$, $P = 0.0003$) followed by a Dunn's correction for multiple comparisons. Each data point represents the mean ± S.E.M. n = 6 rats per group. (**B**) Effect of chronic treatment with JMV2009 (30 µg/kg/day, i.t. for five consecutive days) or Saline (25 µl/day, i.t. for five consecutive days) on the antinociceptive action of JMV2009 (90 µg/kg). The analgesic effect was assessed using the tail-flick thermal test. No statistical differences were found between the anti-nociceptive responses of Saline- or JMV2009-pretreated animals (two-way ANOVA ($F(1, 117) = 1.352$) followed by a Dunnett's post hoc test). Each symbol represents the mean ± S.E.M. of determinations made in 8 rats. (**C**) Hypothermia induced by acute i.t. injection of saline, JMV2009 (3, 30, and 90 µg/kg) and the $NTS_1$-selective compound PD149163 (30 µg/kg). Baseline body temperature (BL) assessment was performed before i.t. injection. As opposed to PD149163, JMV2009 did not induce changes in body temperature (Δ Body Temperature). *** $P < 0.001$ (PD149163, 30 µg/kg vs. Saline) in a two-way ANOVA ($F(4, 231) = 22.19$) followed by a Dunnett's post hoc test. Error bars represent mean ± S.E.M. of determinations made in 8 rats. (**D**) Maximal effect of JMV2009 or NT on mean arterial blood pressure recorded on anesthetized rats after i.v. injection at a dose of 0.1 mg/kg. Each data point represents the mean ± S.E.M. n = 6 rats per group.

# Figures

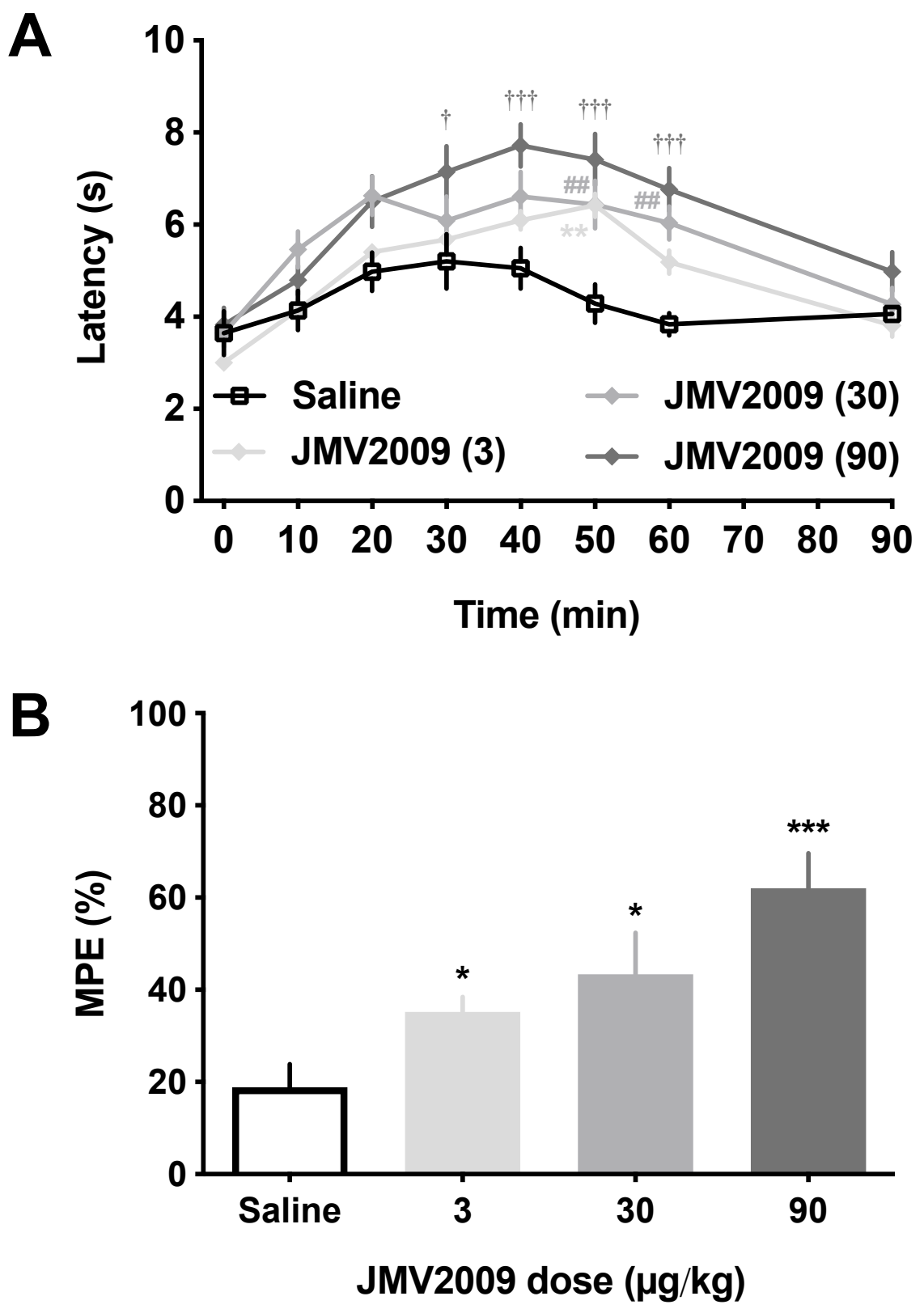

Fig. 1

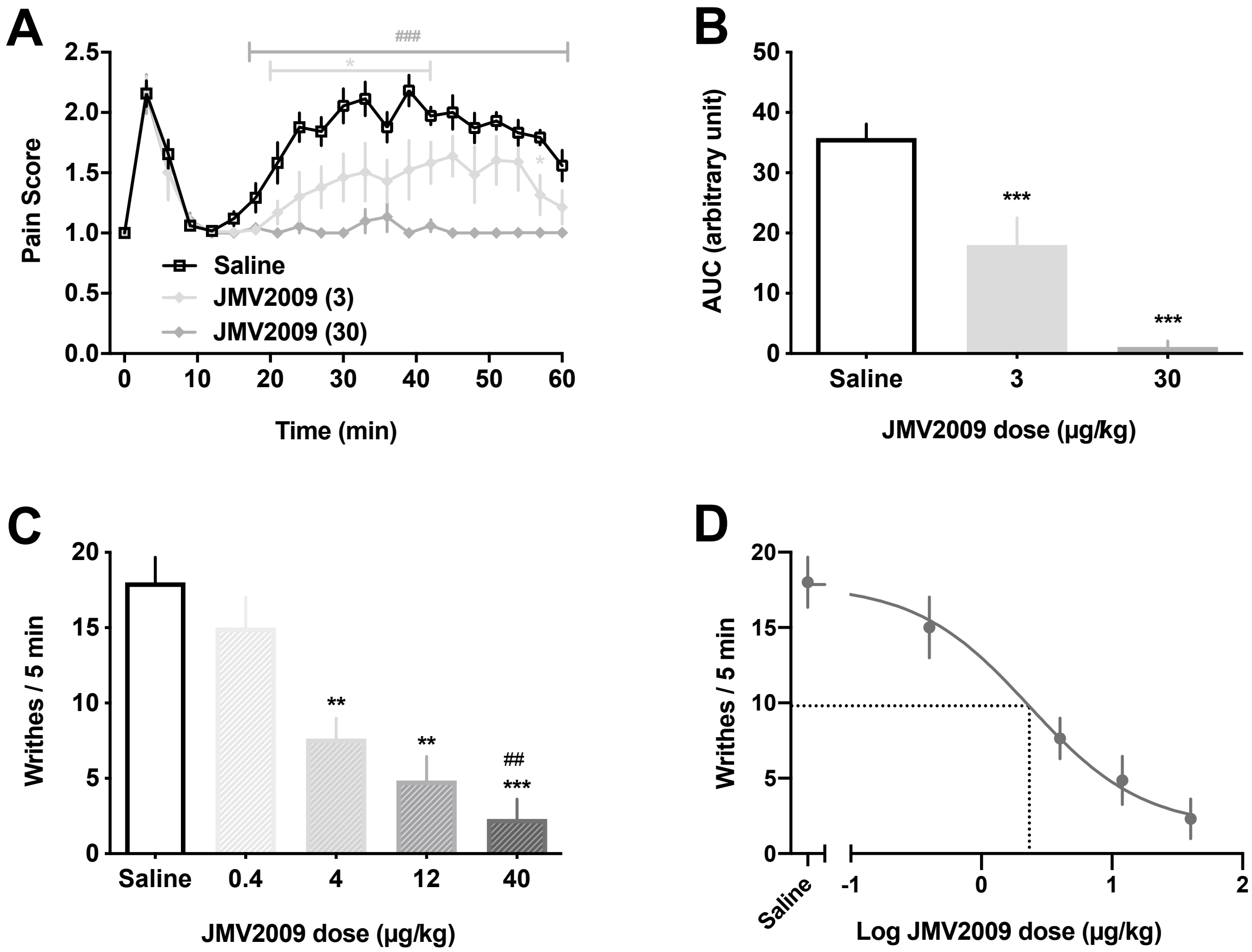

Fig. 2

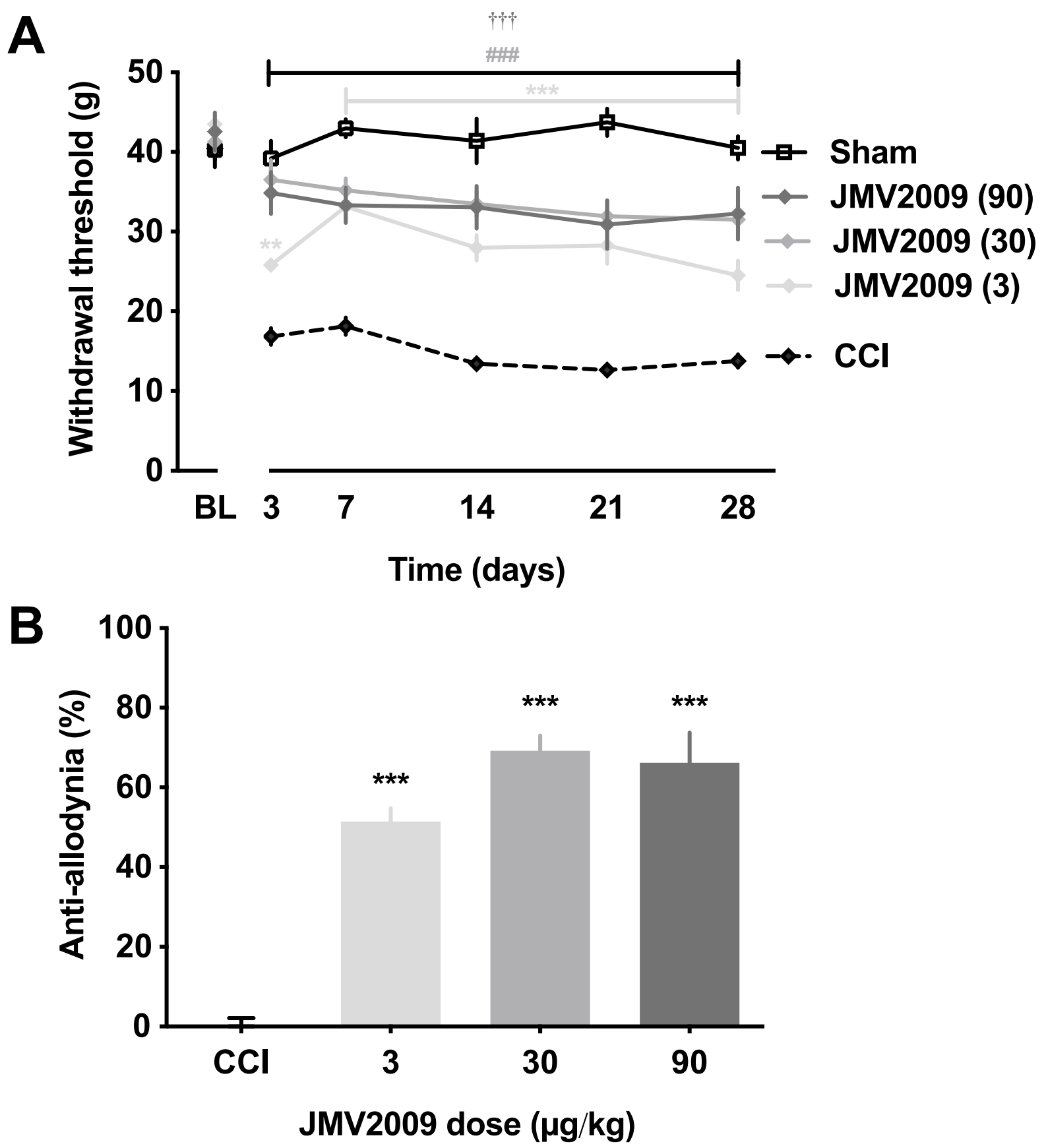

Fig. 3

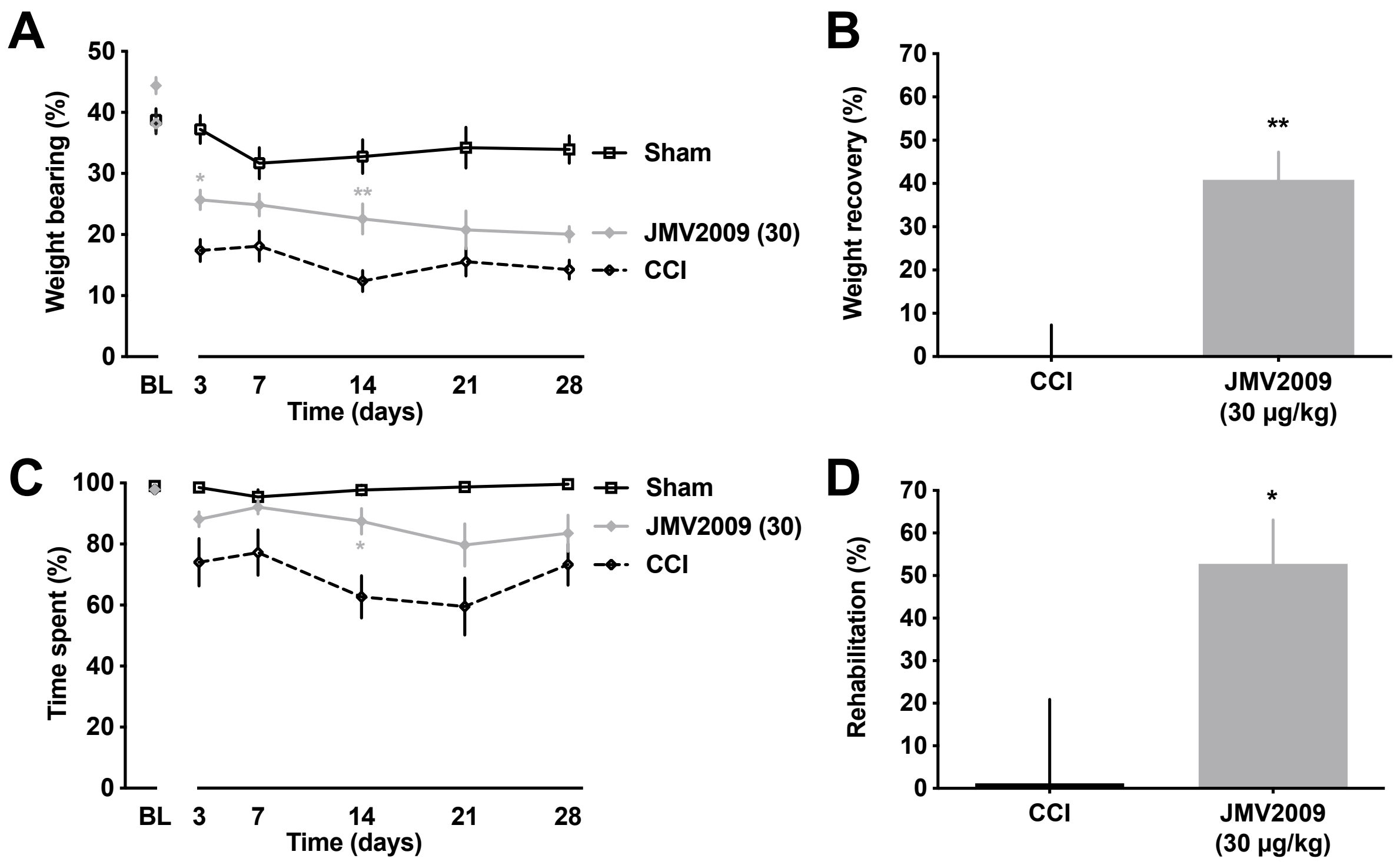

Fig. 4

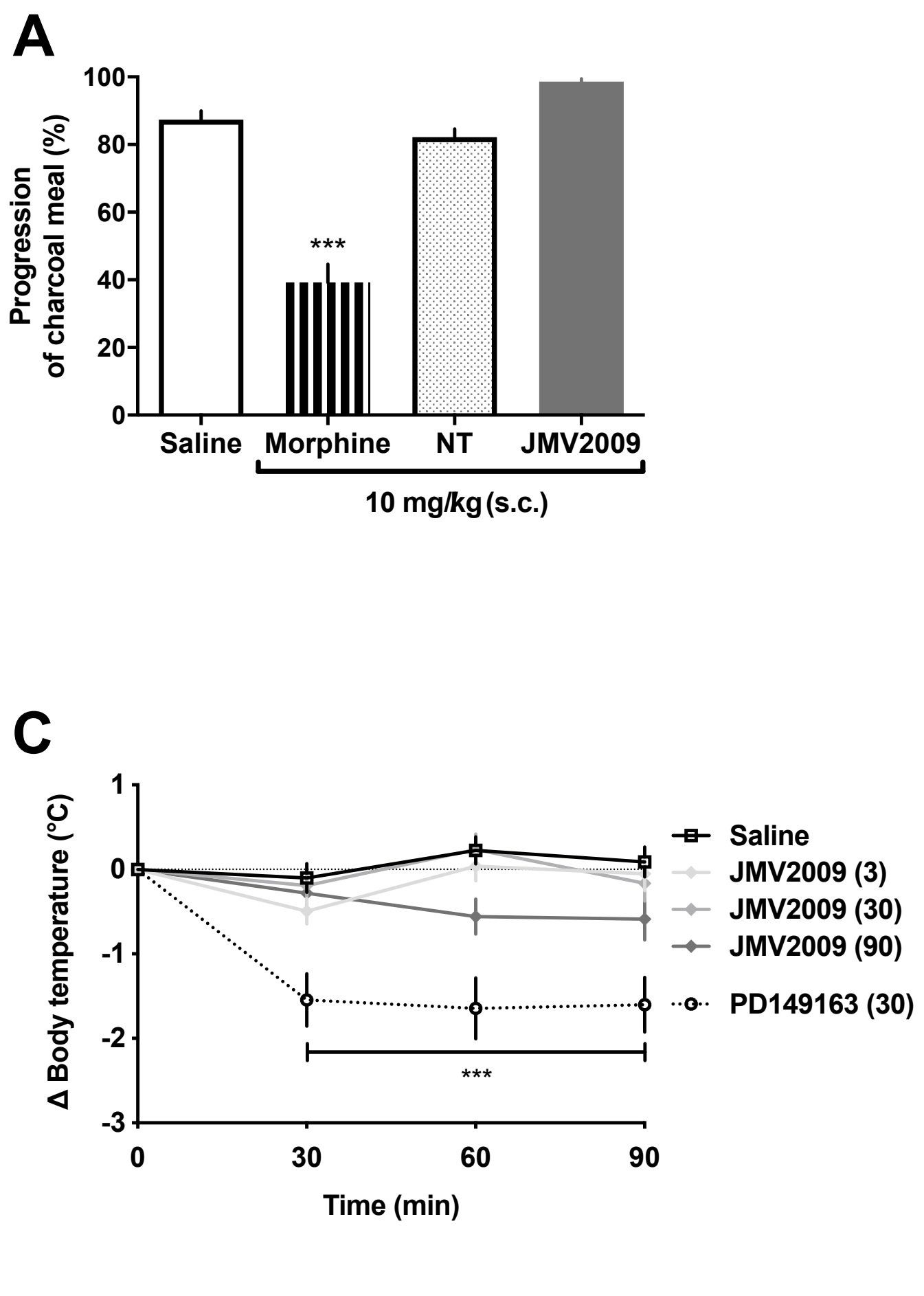

Fig. 5